\documentclass[12pt,preprint]{aastex}

\shorttitle{\emph{XMM-Newton} Observations of the Cataclysmic Variable GW Lib}
\shortauthors{Hilton et al.}

\begin{document}

\title{\emph{XMM-Newton} Observations of the Cataclysmic Variable GW Lib}

\author{Eric J. Hilton, Paula Szkody, Anjum Mukadam}
\affil{Astronomy Department, Box 351580, University of Washington, Seattle WA 98115}
\email{hilton@astro.washington.edu}

\author{Koji Mukai}
\affil{NASA/Goddard Space Flight Center, Greenbelt, MD 20771}
\author{Coel Hellier and Liza van Zyl}
\affil{Astrophysics Group, Keele University, Keele, Staffordshire ST5 5BG, UK}
\author{Lee Homer}
\affil{Liverpool CC, Liverpool, UK}

\begin{abstract}
\emph{XMM-Newton} observations of the accreting, pulsating white dwarf in 
the quiescent dwarf nova GW Librae were conducted to determine if the non-radial 
pulsations present in previous UV and optical data affect the X-ray emission.
The non-radial pulsations are evident in the simultaneous Optical Monitor data but
are not detected in X-ray with an upper limit on the pulsation amplitude of 0.092 mags.
The best fits to the  X-ray spectrum are with a low temperature diffuse gas 
model or a multi-temperature 
cooling flow model, with a  strong OVIII line, similar to other short period 
dwarf novae, but with a 
lower temperature range than evident in normal short period dwarf novae.
The lack of pulsations and the spectrum likely indicate
that the boundary layer
does not extend to the surface of the white dwarf.
\end{abstract}

\keywords{stars: individual -- GW Lib -- stars: dwarf novae -- X-rays: stars}

\section {Introduction}

GW Librae was the first cataclysmic variable in which the white
dwarf was found to exhibit non-radial pulsations \citep{Warner1998}. 
Since that first discovery,
ten other systems have been identified (see recent summary in \citet{Mukadam2007}), but GW Lib remains the brightest and best
studied of the accreting, pulsating white dwarfs. It has a very
short orbital
period of 77 min \citep{Szkody2000,Thorstensen2002}
 and a very low accretion rate, resulting in a
large contribution of the white dwarf to the system light in the
optical and ultraviolet regions of the spectrum. Parallax and proper motion
measurements by \citet{Thorstensen2003} give a distance of 104 pc.
 Until
recently, only one outburst of GW Lib was known \citep{Maza1983} but
in April 2007, a second outburst began. Monitoring of the superhumps
that developed during the course of the outburst \citep{Kato2007}
revealed a period excess of the superhump period over that of the orbital
period of 1.1\%. The relation of period excess to mass ratio and
orbital period \citep{Patterson2001} then implies that the secondary in GW
Lib has passed the period minimum for close binary evolution and is
degenerate.

Photometry over 
several years has shown the characteristics of the pulsations of GW Lib
\citep{vanzyl04}. There are three primary pulsation periods, although
these all show different amplitudes at different times and some of
the periods are not always visible. The most common periods are near
650, 370 and 230 s with typical amplitudes of 0.15, 0.010 and 0.007 mags.
\citet{woudt02} also identified a long period of 2.09 hrs that
was present in 2001 observations but not during 1997-1998. The origin of
this period is unknown but these long periods are present in several
short orbital period disk systems.

Analysis of HST ultraviolet data \citep{Szkody2002} showed the same
pulsations were present in the UV as the optical, but the amplitudes
were about six times larger. An unexpected result from the HST study
was that the best model fit to the spectrum was with a two-temperature
white dwarf, with a T$_{eff}$=13,300K for 63\% of the white dwarf
surface and 17,100K for the remaining 37\%. It was unclear whether the
dual temperatures were a result of the boundary layer (where the
fast moving layers of the inner disk meet the slower rotation of the
white dwarf) providing 
accretion heating of the equatorial regions of the white
dwarf, or due to the pulsations. Further UV studies of accreting
pulsating white dwarfs \citep{Szkody2007} have not shown this
dual temperature structure.

While the soft X-ray emission from the stellar photosphere of the single
hot white dwarf PG1159-036 is known to exhibit similar pulsations modes
as the optical but with 20-30 times the optical amplitudes \citep{Barstow1986},
the interesting question is whether the non-radial
pulsations affect the boundary layer where the X-rays are produced
in CVs. 
In order for theoretical disk instability models to account for the
long interoutburst timescales between dwarf novae outbursts such as
in GW Lib, the
accretion rate has to be very low, requiring very low viscosity and
truncation of the inner accretion disk, possibly by coronal siphons
or from a strong magnetic field on the white dwarf 
\citep{Meyer1994,Warner1996}. Fitting
of spectral energy distributions to models also often invokes a
truncation of the inner disk in order to alleviate excess UV flux
from the models \citep{Linnell2007}. If the inner disk of GW Lib
is truncated, the
X-ray emission should not be pulsed. However, \emph{Chandra}
data on the dwarf nova U Gem \citep{Szkody2002a} shows that the 
boundary layer is close to the white dwarf and moving at low velocity.
If the boundary layer in GW Lib
extends to the white dwarf surface, the X-ray emission may be
modulated at the same periods evident in the UV and optical.
 Thus, the X-ray emission from
GW Lib could provide some constraints on the location and characteristics
of the boundary layer in low accretion rate systems.

Since GW Lib was not detected in the ROSAT All Sky Survey, nor has
any previous X-ray observation, we obtained time on \emph{XMM-Newton} to
obtain light curves and spectra to determine if GW Lib has the
normal hard X-ray emission that is generally present in all low
mass transfer rate, disk-accreting dwarf novae, and if the X-rays
are modulated by the non-radial pulsations evident on its white
dwarf.

\section{Observations and Data Reduction}

\emph{XMM-Newton} observations of GW Lib on August 25-26, 2005 provided simultaneous optical imaging 
from the Optical Monitor \citep[OM;][]{mason01}, and X-ray data from the EPIC pn \citep{struder01}, 
and two MOS detectors \citep{turner01}.
The pn has roughly twice the effective area of either MOS detector.
Because of a low count rate, the Reflection Grating Spectrograph data were not useful.
The X-ray observations lasted approximately 20ks, while the OM consisted of 5 observations of 
approximately 4ks each.
The UT times, length of total observations, and average count rates are listed in Table \ref{tbl:obs}.

The data were reduced using SAS (ver. 7.0.0) following the guidelines from the main 
\emph{XMM-Newton} Web site (Vilspa) and from the NASA/GSFC XMM-Newton Guest Observer Facility ABC Guide (ver. 2.01).
Calibration files are current to August 15, 2006.
The SAS tools were used to create new event list files from the observation data files.
In order to screen out background flaring events, whole-chip light curves for each detector 
were created in the 10-18 keV range and the data were ignored when the count rate was greater than 
2.0 c/s for the pn and greater than 0.6 c/s for each MOS detector.
These background flaring times when the count rate limits were exceeded were nearly identical for all detectors.
The event list files were also screened with the standard canned expressions.
The source aperture was taken to be circular with a radius of 360 pixels for the pn and 320 pixels for 
both MOS detectors in order to maximize the signal-to-noise.  
For the MOS detectors, the source-free background aperture was taken to be an annulus on the central 
chip centered on the source, while for the pn the background was taken to be rectangular regions on 
adjacent chips with similar \emph{Y} locations as the target.
Energies were restricted to the well-calibrated ranges: 0.2 - 15 keV for spectral analysis and 0.1 - 12.5 keV for light curve analysis.
Events were restricted for the pn to singles (pattern = 0) for the spectrum and singles and doubles (pattern $\leq$ 4) for the lightcurve.
For the MOS detectors, up to quadruples (pattern $\leq$ 12) were permitted for both the spectrum and the lightcurve.
FTOOLS\footnote{http://heasarc.gsfc.nasa.gov/ftools/} \citep{Blackburn1995} software tasks were used to 
group the spectral bins and associate various files for spectral analysis in XSPEC, create background 
subtracted light curves, and correct the time stamps to the solar system barycenter.

Data from both MOS detectors and the pn were combined to construct the X-ray 
light curve, which had an average
count rate of 0.042 c/s.
Only data when all three detectors were live and free of background flaring
events were kept.
These are called good time intervals.
Data were also binned to increase the signal-to-noise of this faint source.
The time bin size was chosen to be 150 seconds to simultaneously optimize
signal-to-noise with time
resolution.
Although the time bins were primarily 150 seconds, the time bins at the
edges of the good time intervals
were of different sizes to accommodate all the data.
The time bin size is discussed further in section 3.1.2.

For the OM observations, the B filter was used, and the Pipeline light curves were binned at 50 seconds for
the analysis.
The average count rate for the OM is 6.2 c/s, which is equivalent to a B magnitude of 17.3.

\section{Results}

\subsection{Light Curves}

\subsubsection{Optical}
\begin{figure}
  \epsscale{.80}
  \plotone{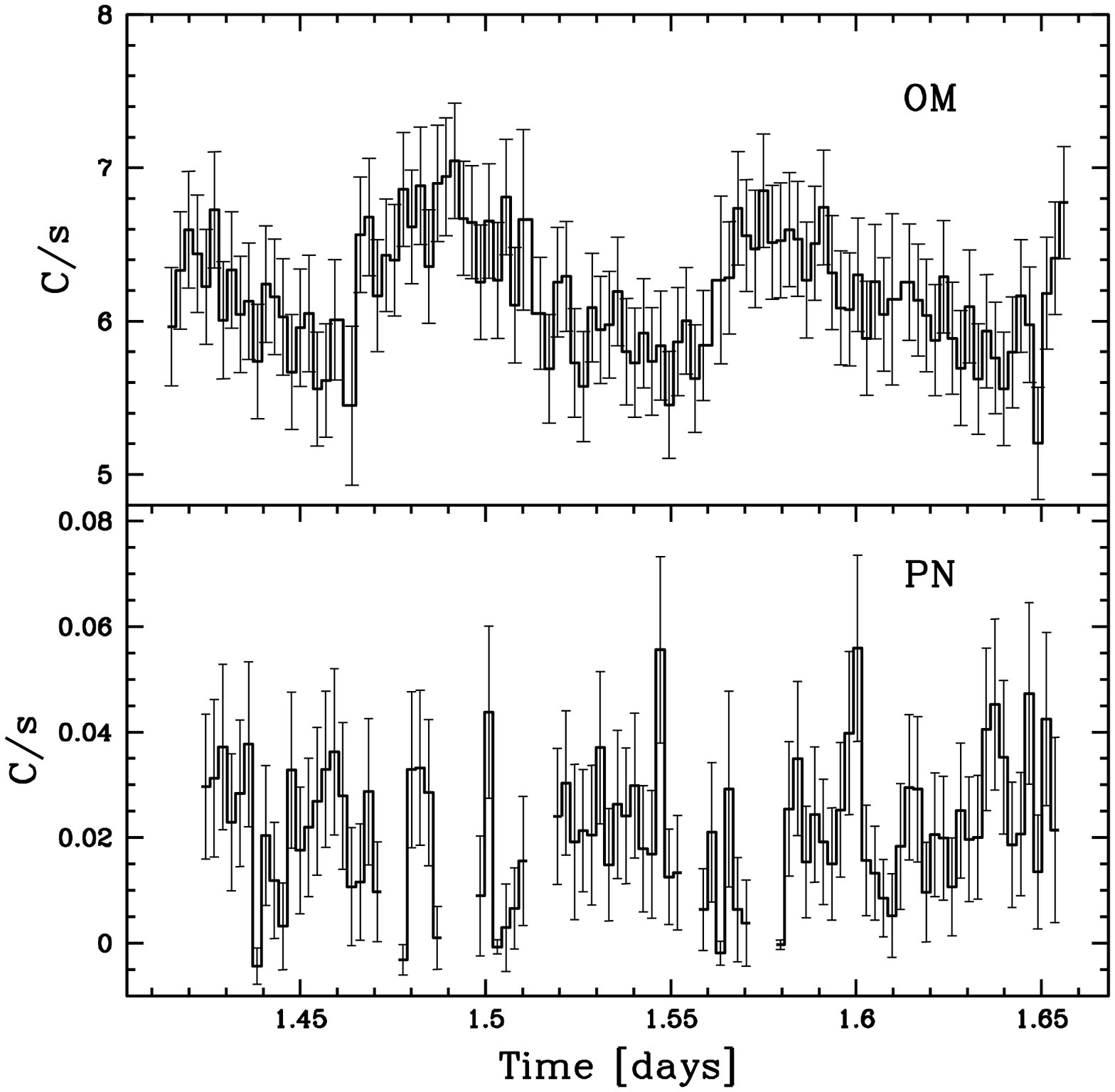}
  \caption{The optical lightcurve (top, shown here binned at 200 sec for clarity) shows a 2.09 hour period.
  The X-ray lightcurve (bottom) shows no apparent period.
  Notice the low count rate of the X-ray data.    
   \label{fig:light_curve} }
\end{figure}

\begin{figure}
  \epsscale{.80}
  \plotone{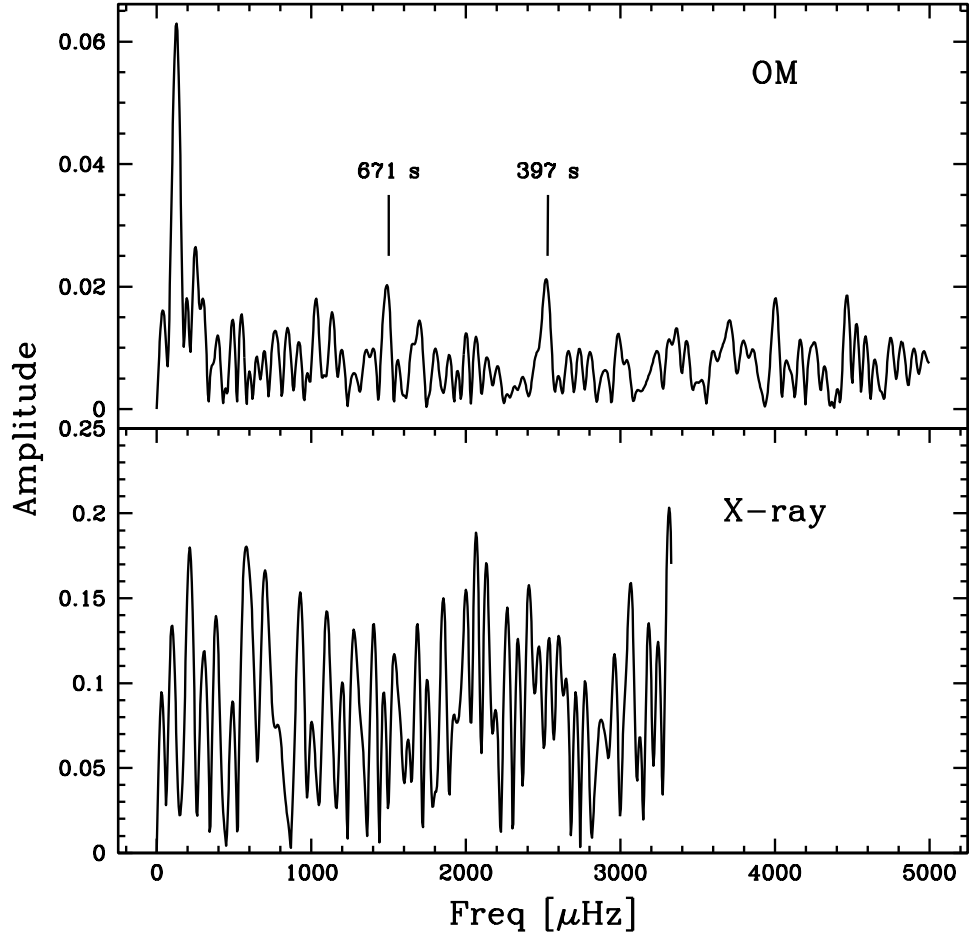}
  \caption{The DFT of the optical lightcurve binned at 50 sec (top) shows the 2.09 hour period as well as 
  the two labelled periods from \citep{vanzyl04}.
  The DFT of X-ray light curve binned at 150 sec (bottom) shows no periods.
  Note that with the longer time bins for the X-ray data, the Nyquist frequency is lower, and the DFT 
  does not extend to as high frequencies as the OM data.
  See text for a discussion of the limit of variability.  \label{fig:dft} }
\end{figure}

The optical light curve of GW Lib, shown in Figure \ref{fig:light_curve}, is dominated by the 2.09 hour period
that was intermittently present in the data of  \citet{woudt02}.
The discrete Fourier transform (DFT) of the optical data shown in Figure \ref{fig:dft} 
shows this long period as well as modulations at 671 seconds with an amplitude of 0.02 mags and 397 seconds 
with an amplitude of 0.021 mags.
These modulations are consistent with the previously observed pulsation periods near 650 s (1540 $\mu$Hz) 
and 370 s (2700 $\mu$Hz), whose periods and amplitudes are known to vary \citep{vanzyl04}.
Van Zyl et al. also find a pulsation near 230 s (4350 $\mu$Hz) that is not seen in the OM data. 
However, the typical amplitude of this period is below the average noise level of this DFT, so its 
presence cannot 
be ruled out.

\subsubsection{X-ray}

The DFT of the combined X-ray data showed no significant periodicities.
In order to place an upper-limit on the magnitude of variability, the
following light-curve shuffling
technique was applied to empirically determine the noise in the light
curve.
A light curve consists of a series of fractional intensity values each with
a corresponding time value.
Each value of fractional intensity was randomly reassigned to one of the
unchanged, existing time values.
This random shuffling destroys any coherent frequencies in the light curve
but maintains the same time
sampling and random white noise as the original light curve.
The DFT of the shuffled light curve gives the amplitude of the noise at each
frequency up to the Nyquist frequency.
The original light curve was randomly shuffled 10 times and the average
noise was computed each time.
The noise of the original light curve was taken to be the mean of these 10
values.

As a check on the time bin size, light curves were produced with time bins
of  primarily 50, 75, 100,
150, and 200 seconds.
In all cases, there were no strong signals present in the light curves and
there were no significant
 differences in the average noise values.
Because the count rate was so low, the time bin size was chosen to maximize
the signal-to-noise without
destroying the time resolution.
Since the shortest period seen in the simultaneous optical observations was
397 seconds, the 150 second
time resolution provides more than two points per cycle, which is
sufficient time resolution.
The unshuffled DFT is shown in Figure \ref{fig:dft}.
The average noise averaged over ten random shufflings is 0.092 mags, which
is taken to be the upper
limit of the X-ray
pulsations for GW Lib.

\subsection{Spectral Analysis}

The extracted background-subtracted spectrum from the pn detector was 
binned at 10 counts per bin to 
facilitate the use of $\chi^2$ statistics to find the best fit models.
The spectrum was restricted to the energy range 0.2-15.0 keV because the 
calibration of the EPIC detectors 
at the lowest energies is not certain and the count rate above 15.0 keV is too low to be useful.
Although the data reduction allows high energy photons, there were very few photons detected with energies 
greater than 3 keV.
The spectrum has a strong \ion{O}{8} emission line at $\sim$0.65 keV and 
an increase in emission at 
$\sim$ 1.0 keV that is possibly a Ne-Fe emission complex.
Several models were used, starting with the simplest emission mechanisms 
(bremsstrahlung), and advancing in 
complexity to more detailed models and variable abundances. 
All models used absorption, but since all models 
consistently found a low value for the hydrogen column density, 
it was subsequently fixed at $10^{20}$ cm$^{-2}$ to reduce the number of 
parameters.  
The redshift was fixed at $10^{-9}$ for the \textbf{mekal} and \textbf{mkcflow}
 families of models and the hydrogen density of the gas was fixed at 
0.1 cm$^{-3}$ for the \textbf{mekal} family of models.
Parameters of the model and the goodness of fit statistics are listed in Table \ref{tbl:fits}.

The simple absorbed bremsstrahlung (\textbf{wabs(bremss)}) model had a reduced $\chi^2 = 1.05$, but was 
unable to fit the strong emission lines.
Explicitly adding a Gaussian to model the oxygen line decreased the 
residuals, and had a reduced $\chi^2 = 0.73$, but was
unable to fit the lines near 1 keV.

\begin{figure}
  \epsscale{.80}
  \includegraphics[angle=270,scale=.60]{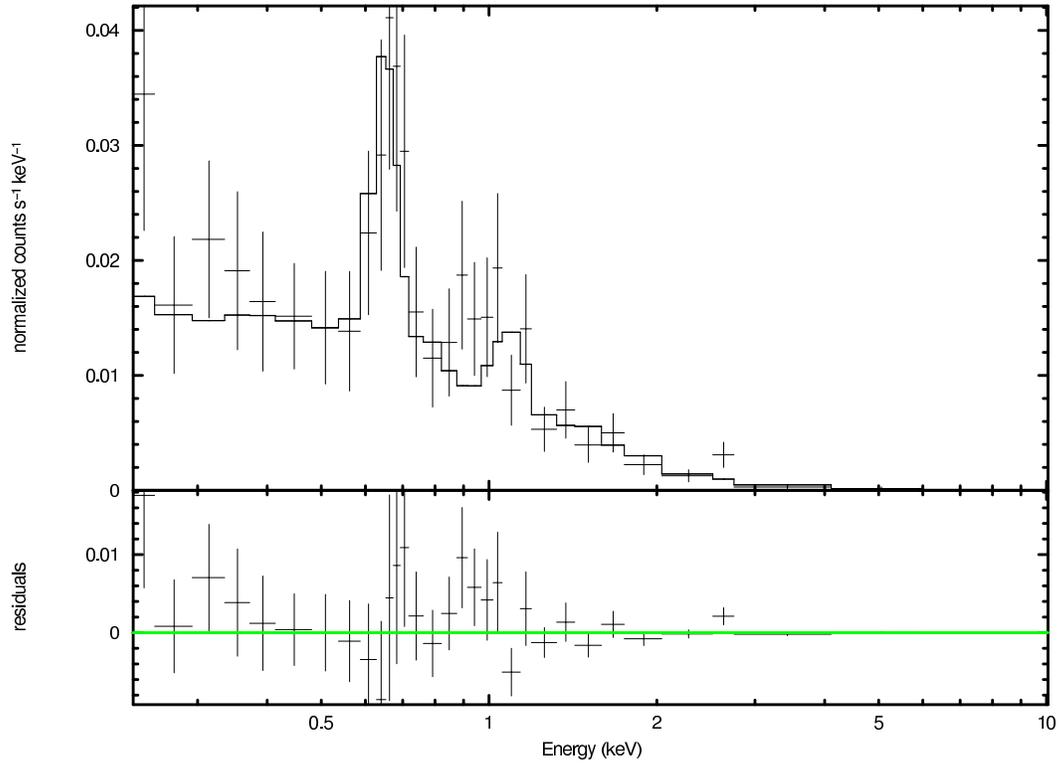}
  \caption{The hot diffuse gas model with variable oxygen abundance.
  \label{fig:vmekal} }
\end{figure}

The model of hot diffuse gas with line emissions from several elements \textbf{wabs(mekal)} with a solar 
abundance mixture also could not fit the emission lines (reduced $\chi^2 = 0.95$).
The variable abundance version of this model (\textbf{wabs(vmekal)})  
gave a better fit to the both the $\sim$0.65 keV and the 
$\sim$1.0 keV lines.
All combinations of 
varying the oxygen, neon, and iron abundances were tried.
As there were no significant differences in the model fits with different 
iron and neon abundances, these were finally left fixed at solar abundance.
The model with oxygen as a parameter of the fit is shown in 
Figure \ref{fig:vmekal}, and has a reduced $\chi^2 =0.81$.

\begin{figure}
  \epsscale{.80}
  \includegraphics[angle=270,scale=.60]{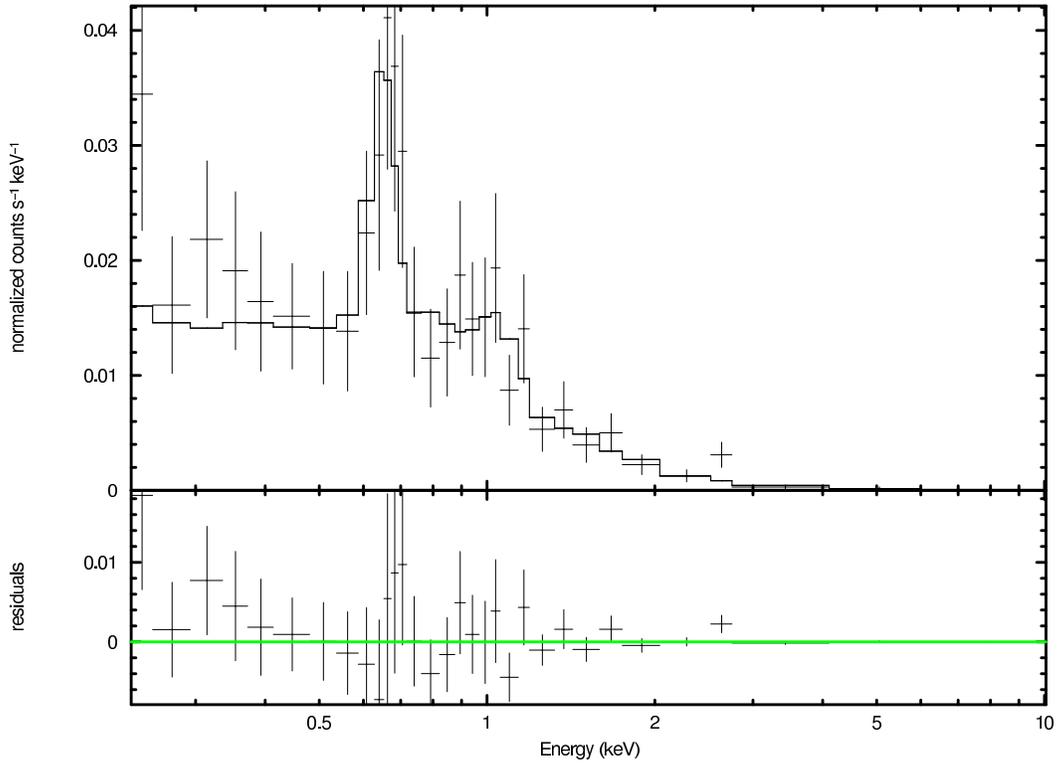}
  \caption{The cooling flow model with variable oxygen abundance.
  \label{fig:vmcflow} }
\end{figure}

\citet{Mukai2003} and \citet{Pandel2003} found successful fits using a cooling flow model 
(\textbf{wabs(mkcflow)}), so this model was also tried, although it 
did not fit the oxygen line nor fully fit the lines at $\sim$1.0 keV.
Adjusting the oxygen and neon abundances using (\textbf{wabs(vmcflow)}) did give a better fit to the 
emission lines with significantly higher oxygen abundance (compared to solar) and a slightly increased 
neon abundance.
Since the mekal models showed that the neon abundance was very uncertain, the cooling flow model was also 
tried with leaving the neon fixed at solar abundance and allowing only the oxygen to be fit.
This model is shown in Figure \ref{fig:vmcflow}.

There are still residuals in both the vmekal and vmcflow model fits (Figures \ref{fig:vmekal} and \ref{fig:vmcflow}) near 0.9 keV.
A Gaussian was added to the vmekal model at that energy but there was no significant
improvement in the fits.
Regardless of the model that was fit to the data, the temperature is generally low (1.5 - 2.5 keV) 
compared to most dwarf novae \citep{Ramsay2001,Pandel2003,Hakala2004}.

\section{Discussion}

The X-ray flux of GW Lib is much lower than expected for its optical magnitude and physical parameters.
The cataclysmic variable WZ Sge has an orbital period and long-term outburst 
characteristics similar to GW Lib. The absolute visual magnitudes of the two systems 
are comparable
(11.8 for WZ Sge and 11.9 for GW Lib) and the white dwarfs have comparable temperatures.
Using the 4.5 keV thermal bremsstrahlung model of WZ Sge \citep{Patterson1998} as a comparison, and correcting for 
distance, PIMMS predicts a count rate of about 0.1-0.2 c/s for GW Lib with 
the EPIC pn 
detector.
The actual average count rate was much lower: 0.02 c/s for the pn and only 0.04 c/s after 
combining all three X-ray detectors (see Figure \ref{fig:light_curve}).
\emph{XMM-Newton} observations of other relatively nearby short period 
dwarf novae (T Leo, OY Car, VW Hyi, WX Hyi, SU UMa, TY PsA and YZ Cnc with
orbital periods between 85-125 min and more frequent outbursts than GW Lib and
WZ Sge), the pn count 
rates were between 1-7 c/s \citep{Ramsay2001,Pandel2003,Hakala2004,Pandel2005}
The 0.2-10 keV fluxes for the best fit vmekal (hot diffuse gas) and vmcflow 
(cooling flow) models for GW Lib shown in Figures \ref{fig:vmekal} 
and \ref{fig:vmcflow} are 6.82 $\times 10^{-14}$ and 6.90 $\times 10^{-14}$ ergs cm$^{-2}$ s$^{-1}$ 
respectively. For a distance of 104 pc \citep{Thorstensen2003}, the X-ray luminosity would be
9$\times 10^{28}$ ergs s$^{-1}$. This compares to L$_{x}$ of 4$\times 10^{30}$, 
8$\times 10^{30}$ and
1.4$\times 10^{32}$ ergs s$^{-1}$ for OY Car, VW Hyi and YZ Cnc. Assuming
this is the boundary layer luminosity, and using the relation given in
\citet{Pandel2003}:

L$_{bl}$ = 5/2 kT$_{max}$/$\mu$ m$_{p}$

where T$_{max}$ is the maximum temperature in the cooling flow model
(5 keV), $\mu$ =0.6 and m$_{p}$ is the proton mass, we can estimate
that $\dot{M}_{bl}$ = 7$\times 10^{-14}$ M$_{\odot}$ yr$^{-1}$. This value
is typically 2 orders of magnitude lower than that for the other dwarf
novae \citep{Pandel2005}. This value is also much lower than the time-averaged
$\dot{M}$ of 7.3$\times 10^{-11}$ estimated by \citep{Townsley2004} from
their model parameters for GW Lib. 

All model fits to the spectrum of GW Lib resulted in
 lower temperatures compared to OY Car, VW Hyi and YZ Cnc and the other systems
as well. Although the best fit 
to all systems involve a range of temperatures, the
maximum temperature for GW Lib is around 5 keV while the kT$_{max}$ for
the short period objects in Pandel et al. ranges from 8-26 keV. The low temperature
is likely not due to an exceptionally low mass for the primary in GW Lib, as
the UV fits \citep{Szkody2002} and the pulsation models \citep{Townsley2004} indicate
a high mass white dwarf.  
The low temperature in GW Lib suggests that the accreting gas is low density or is 
only mildly shocked, so the X-ray cooling is very inefficient. It is likely that the
shock occurs high above the white dwarf surface which lowers the shock temperature.
The stronger oxygen line in GW Lib compared to these other systems and the
lack of FeK$\alpha$ at 6.4 keV are likely  artifacts of the low temperature (although
we cannot rule out that there is some peculiar atomic physics that is not taken
into account in the mekal-type models).
It is noteworthy that FeK is also missing in WZ Sge and its temperature is
similar to GW Lib \citep{Patterson1998} so the lower accretion rates in
these systems with rare but tremendous amplitude outbursts \citep{Howell1995} likely lead to
similar weak boundary layers.
 
The pulsations that are visible in the optical and UV are limited to
an X-ray amplitude of $<$ 0.09 mag. The low X-ray flux, cool temperatures
and absence of
strong X-ray pulsations all imply that the boundary layer in GW Lib
does not reach to
the white dwarf surface to create a strong shock or to be affected by
the surface pulsations.

The lack of X-ray modulation at the 2.09 hr period argues against an
origin for this period in the inner disk of a magnetic, precessing
white dwarf as has been suggested for the long periods seen in FS Aur
and HS2331+39 \citep{Tovmassian2007}.

\section{Conclusions}

The \emph{XMM-Newton} observations of GW Lib have shown that the X-ray
emitting region of the accreting, pulsating white dwarf is not
strongly affected by the non-radial pulsations evident in the UV and optical. 
The unusually weak
X-ray flux from this system precludes a stringent limit, but does rule
out pulsation amplitudes of greater than 0.09 mag, specifically at the periods where 
significant signals are detected simultaneously in the optical band.
The low X-ray flux, cool maximum temperature of the X-ray spectrum,
combined with the lack of X-ray pulsation indicate the boundary layer
is not dense enough to create a strong shock at the white dwarf surface.
This has implications for the two-temperature model for the white
dwarf that was needed to explain the HST UV spectrum \citep{Szkody2002}
in that the origin of the hotter temperature component
 may be related to the pulsations, and
not to boundary layer heating. 

\acknowledgements

This work was supported by \emph{XMM-Newton} grant NNG05GR47G to the University
of Washington and is based on observations obtained with \emph{XMM-Newton},
an ESA science mission with instruments and contributions directly funded
by ESA Member States and the USA (NASA).


\clearpage

\clearpage

\clearpage

\clearpage

\clearpage
\begin{deluxetable}{llllll} 
 \tabletypesize{\scriptsize}
\tablewidth{0pt}
\tablecaption{25-26 August, 2005 Observations.\label{tbl:obs}}
\tablehead{
\colhead{Instrument} & \colhead{Filter} & \colhead{Duration (s)} &\colhead{UT Start Time} &\colhead{UT Stop Time} &\colhead{Ave. count rate (C/s)\tablenotemark{a}}}
\startdata
PN   & Thin1 & 19936 & 22:09:28  &  03:41:44 & (2.32$\pm$ 0.15)$\times10^{-2}$ \\
MOS1 & Thin1 & 21809 & 21:47:09  &  03:50:38 & (7.06$\pm$ 0.72)$\times10^{-3}$ \\
MOS2 & Thin1 & 21577 & 21:47:09  &  03:46:46 & (8.43$\pm$ 0.78)$\times10^{-3}$ \\
OM   & B     & 19901 & 21:55:31  &  03:48:49 & 6.2 $\pm$ 0.6 (B = 17.3)\tablenotemark{b} \\
\enddata
\tablenotetext{a}{X-ray count rates determined from spectral reductions}
\tablenotetext{b}{OM count rate determined from light curve and converted to standard B magnitude}
\end{deluxetable}

\clearpage
\begin{deluxetable}{lllll} 
\tabletypesize{\scriptsize}
\tablewidth{0pt}
\tablecaption{XSPEC models used to fit the X-ray spectrum.\label{tbl:fits}}
\tablehead{
\colhead{Model Name} & \colhead{Reduced $\chi^2$} & \colhead{kT} &\colhead{Normalization} &\colhead{Parameters}}
\startdata

Bremss\tablenotemark{a}         & 1.05   & 2.2  & 2.3$\times 10^{-5}$ &  \\

Bremss\tablenotemark{b} & 0.73   & 2.11 & 2.1 $\times 10^{-5}$&  LineE = 0.67 keV \\
+ Gauss  &&&&$\sigma$ = 9.9$\times 10^{-7}$ keV\\
&&&& norm = 4.1$\times 10^{-6}$\\

mekal\tablenotemark{c}         & 0.95  & 1.90 &  6.0$\times 10^{-5}$  & abundance = 0.19   \\

vmekal\tablenotemark{d} & 0.81 & 2.50 & 2.9$\times10^{-5}$ & O abund. = 7.99$\times$solar \\

mkcflow\tablenotemark{e} &   0.83   &  0.091 - 4.70 & 1.6$\times 10^{-16}$ & Abundance = 0.28   \\

vmcflow\tablenotemark{f}  &  0.70 & 0.38 - 5.52  & 1.4$\times 10^{-16}$ & O abund. = 6.3$\times$solar \\
&&&& Ne abund. = 1.4$\times$solar \\

vmcflow\tablenotemark{g} & 0.67 & 0.38 - 5.45 & 1.5$\times10^{-16}$ & O abund. = 6.12$\times$solar \\

\enddata
\tablenotetext{a}{Thermal Bremsstrahlung - Didn't fit the emission lines }
\tablenotetext{b}{Thermal Bremsstrahlung plus Gaussian -  Fit the oxygen line well}
\tablenotetext{c}{Emission from a hot diffuse gas - model shows bump at 0.6 keV, but doesn't fit line}
\tablenotetext{d}{Emission from a hot diffuse gas with variable abundances, neon fixed at solar}
\tablenotetext{e}{Cooling flow}
\tablenotetext{f}{Cooling flow with variable abundances}
\tablenotetext{g}{Cooling flow with variable abundances, neon fixed at solar}

\end{deluxetable}

\end{document}